\documentclass{aa}
\usepackage{graphicx}

\newcommand{\vd}{\overrightarrow{d}}
\newcommand{\vTeta}{\overrightarrow{\Theta}}
\newcommand{\mC}{\vec{C}}
\newcommand{\mT}{\vec{T}}
\newcommand{\mN}{\vec{N}}
\newcommand{\mA}{\vec{A}}
\newcommand{\mW}{\vec{W}}
\newcommand{\Npix}{N_{\rm pix}}
\newcommand{\fbP}{\delta T_{\rm fb}}
\newcommand{\EfbP}{\hat{\delta T_{\rm fb}}}
\newcommand{\vfbP}{\overrightarrow{\delta T_{\rm fb}}}
\newcommand{\sigM}{\sigma_M}
\newcommand{\EsigM}{\hat{\sigma}_M}
\newcommand{\sigN}{\sigma_N}

\begin{document}

   \thesaurus{12.03.1;12.03.3;12.03.4;12.04.2;12.12.1;11.03.1} 
   \title{An Approximation to the Likelihood Function for Band--Power
        Estimates of CMB Anisotropies}

   \subtitle{}

        \titlerunning{An Approximation...} 

   \author{J.G.~Bartlett$^1$, M.~Douspis$^1$, A.~Blanchard$^{1,2}$ 
        \& M.~Le~Dour$^1$}

   \offprints{bartlett@ast.obs-mip.fr}

   \institute{$^1$ Observatoire Midi-Pyr\'en\'ees,
              14, ave. E. Belin,
              31400 Toulouse, FRANCE \\
              Unit\'e associ\'ee au CNRS 
             ({\tt http://www.omp.obs-mip.fr/omp})\\    
              $^2$ Observatoire de Strasbourg,
                   Universit\'e Louis Pasteur,
                   11, rue de l'Universit\'e,
                   67000 Strasbourg,
                   FRANCE\\
                   Unit\'e associ\'ee au CNRS 
                   ({\tt http://astro.u-strasbg.fr/Obs.html})
             }

   \date{June 29, 1999}

   \maketitle

   \begin{abstract}
        Band--power estimates of cosmic microwave background
fluctuations are now routinely used to place constraints on
cosmological parameters.  For this to be done in a 
rigorous fashion, the full likelihood function of 
band--power estimates must be employed.  Even for Gaussian
theories, this likelihood function is not 
itself Gaussian, for the simple reason that band--powers 
measure the {\em variance} of the random sky 
fluctuations.  In the context of Gaussian sky
fluctuations, we use an ideal situation to motivate 
a general form for the full likelihood function from a given
experiment.  This form contains only two free parameters, which
can be determined if the 68\% and 95\% confidence
intervals of the true likelihood function are known.    
The ansatz works remarkably well when compared to the 
complete likelihood function for a number of experiments.
For application of this kind of approach, we suggest 
that in the future both 68\% and 95\% 
(and perhaps also the 99.7\%) confidence intervals be
given when reporting experimental results.
   \end{abstract}


\section{Introduction}

        Six years after their first detection
by the COBE satellite (Smoot et al. 1992), it 
is now well appreciated that cosmic microwave
background (CMB) temperature fluctuations contain rich 
information concerning virtually all the fundamental cosmological
parameters of the Big Bang model (Bond et al. 1994;
Knox 1995; Jungman et al. 1996).  
New observations from 
a variety of experiments, ground--based and balloon--borne,
as well as the two planned satellite missions, MAP\footnote{
{\tt http://map.gsfc.nasa.gov/}} and Planck 
Surveyor\footnote{{\tt http://astro.estec.esa.nl/Planck/}},
are and will be supplying a constant stream of ever
more precise data over the next decade.

        It is in fact already
possible to extract interesting information from the 
existing data set, consisting of almost 20 different 
experimental results (Lineweaver
et al. 1997; Bartlett et al. 1998a,b; Bond \& Jaffe 1998;
Efstathiou et al. 1998; Hancock et al. 1998; 
Lahav \& Bridle 1998;
Lineweaver \& Barbosa 1998a,b; Lineweaver 1998;  
Webster et al. 1998; Lasenby et al. 1999).
These experimental results are most often given in the
literature as power estimates within a band defined over a 
restricted range of spherical harmonic orders.  Our compilation,
similar to those of Lineweaver et al. (1997) and Hancock et al. (1998),
is shown in Figure 1 and may be accessed at our web 
site\footnote{{\tt http://astro.u-strasbg.fr/Obs/COSMO/CMB/}}.
The band is defined either directly by the observing 
strategy, or during the data analysis, e.g., the electronic
differencing scheme introduced by Netterfield et al. (1997).  
This permits a concise representation 
of a set of observations, reducing a large number of 
pixel values to only a few band--power estimates, and for this
reason the procedure has been referred to as
``radical compression'' (Bond et al. 1998).
If the sky fluctuations are Gaussian, as predicted
by inflationary models, then little or nothing
has been lost by the reduction to band--powers
(Tegmark 1997).
This is extremely important, because the limiting
factor in statistical analysis of the next 
generation of experiments, such as, e.g., 
BOOMERanG\footnote{{\tt http://astro.caltech.edu/$\sim$lgg/boom/boom.html}}, 
MAXIMA\footnote{{\tt http://cfpa.berkeley.edu/group/cmb/gen.html}}, 
and Archeops\footnote{{\tt http://www-crtbt.polycnrs-gre.fr/archeops/general.html}}, 
is calculation time.  Working with 
a much smaller number of band--powers, instead of the
original pixel values, will be essential for 
such large data sets.  
The question then
becomes how to correctly treat the statistical 
problem of parameter constraints starting directly with
band--power estimates.  
\begin{figure}\label{fig_fignew}
\begin{center}
\resizebox{\hsize}{!}{\includegraphics[angle=-90,totalheight=9cm,
        width=8.9cm]{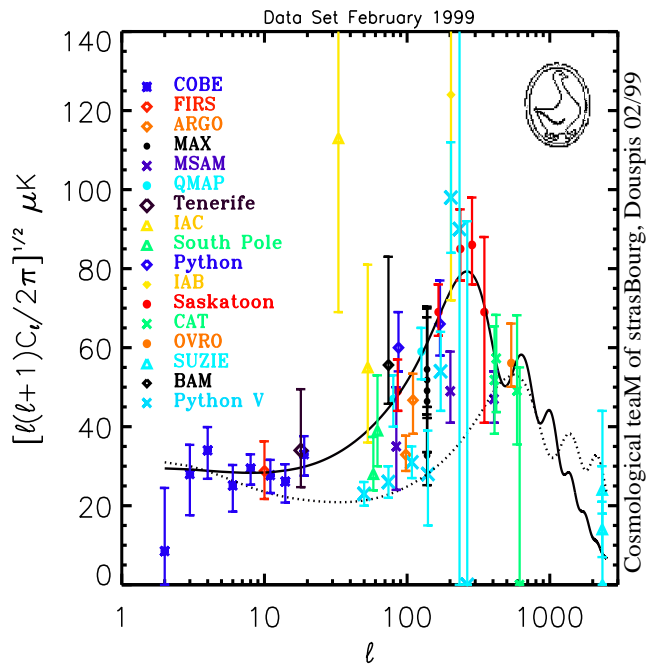}}
\end{center}
\caption{Present CMB power spectrum estimates.  Flat band--powers
are shown as a function of multipole order $l$.  The data 
and references can be found at 
{\tt http://astro.u-strasbg.fr/Obs/COSMO/CMB/tabcmb.html}.
The solid curve is a flat CDM 
model with $\Omega=0.4$, $\Lambda=0.6$, $H_o=40$ km/s/Mpc, $Q=19\;\mu$K, 
$\Omega_b h^2 = 0.006$ and $n=0.94$,
while the dotted line represents an open model with $\Omega=0.2$, 
$\Lambda=0$, $H_o=60$ km/s/Mpc, $Q=20\;\mu$K, $\Omega_b h^2 = 0.015$ 
and $n=1$ (no gravitational waves and no reionization).
}
\end{figure}

        Standard approaches to parameter determination,
whether they be frequentist or Bayesian, begin with 
the construction of a likelihood function.  For Gaussian
fluctuations, the only kind we consider here,
this is a multivariant Gaussian in the pixel temperature 
values, where the covariance matrix is a function 
of the model parameters (see below).  The likelihood
is then used as a function of the parameters, but
as just mentioned, the large number of pixels makes
this object very computationally cumbersome.  
It would be extremely useful to be able to define 
a likelihood function starting directly with the
power estimates in Figure 1.  This is 
the concern of this {\em paper}, where we 
develop an approximation to the the full likelihood
function which requires only band--power 
estimates and very limited experimental details.
As always in such procedures, it is worth
emphasizing that the likelihood function, and
therefore all derived constraints, only applies
within the context of the particular model
adopted.  In our discussion, we shall focus primarily 
on inflationary scenarios, whose 
theoretical predictions have become easily calculable
thanks to the development of fast Boltzmann codes, such
as CMBFAST (Seljak \& Zaldarriaga 1996; Zaldarriaga et al. 1998).

        Much of the recent work
on parameter determination has relied on the traditional
$\chi^2$--fitting technique.  As is well known, this 
amounts to a likelihood approach for observables 
with a Gaussian probability distribution.  Band--power
estimates do not fall into this category (Knox 1995;
Bartlett et al. 1998c; Bond et al. 1998; Wandelt et al. 1998) -- 
they are not
Gaussian distributed variables, not even in the case
of underlying Gaussian temperature fluctuations.  The 
reason is clear: power estimates represent  
the {\em variance} of Gaussian distributed pixel values
(the sky temperature fluctuations), and they therefore have
a distribution more closely related to the $\chi^2$--distribution.

        We begin, in the following section, 
by a general discussion of the likelihood approach 
applied to CMB observations.  In the context of an 
ideally simple situation, we find the {\em exact}
analytic form for the likelihood function of a band--power
estimate.  Reflections concerning the likelihood function 
in the context defined by actual experiments 
motivates us to propose this analytic form as an approximation,
or ansatz, in the more general case.  
It is extremely easy to use, requiring little information 
in order to be applied to an experimental setup, because it 
contains only two adjustable parameters.  
These can be completely determined if one is given
two confidence intervals, say the 68\% and
95\% confidence intervals, of the true, underlying
likelihood distribution (notice that here
we see the non--Gaussian nature of the likelihood --
a Gaussian function would only require one confidence 
interval, besides the best power estimate, to be
completely determined).  We
ask that in the future at least two confidence
intervals be given when reporting experimental
band--power estimates (more would be better, say
for adjusting more complicated functional forms).
An important limitation of the approach is the
inability at present to account for more than one, correlated
band--powers, as will be discussed further below. 

         We quantitatively test the accuracy of the 
approximation in Section 3 by comparison to several
experiments for which we have calculated the full 
likelihood function.  The approximation  
works remarkably well, and it can represent a substantial
improvement over both single and ``2--winged'' Gaussian 
forms commonly used in standard $\chi^2$--analyses;
and it is as easy to use as the latter.
The proposed likelihood approximation, the main
result of this {\em paper}, is given in 
Eqs. (\ref{eq:approx}) -- (\ref{eq:ninty5}).  We plan to 
maintain a web 
page\footnote{{\tt http://astro.u-strasbg.fr/Obs/COSMO/CMB/}} 
with a table of the 
best fit parameters required for its use.
Detailed application
of the approximate likelihood function to parameter 
constraints and to tests of the Gaussianity of the
observed fluctuations is left to future papers.
Other, similar work has been performed by Bond et al. (1998) and
Wandelt et al. (1998).

\section{Likelihood Method}

\subsection{Generalities}

        Temperature anisotropies are 
described by a 2--dimensional {\em random}
field $\Delta(\hat{n})\equiv (\delta T/T) (\hat{n})$, 
where $\hat{n}$ is
a unit vector on the sphere. This means 
we imagine that the 
temperature at each point has been randomly
selected from an underlying probability 
distribution, characteristic of 
the mechanism generating the perturbations
(e.g., Inflation).
It is convenient to expand the field in spherical 
harmonics:
\begin{equation}
\Delta(\hat{n}) = \sum_{lm} a_{lm} Y_{lm}(\hat{n})
\end{equation}
For Inflation generated perturbations, 
the coefficients $a_{lm}$ 
are {\em Gaussian random variables} with zero mean  --
$<a_{lm}>_{ens} = 0$ -- and covariance 
\begin{equation}\label{eq:defCl}
<a_{lm}a^*_{l'm'}>_{ens} = C_l \delta_{ll'}\delta_{mm'}
\end{equation}
This latter equation defines
the {\em power spectrum} as the set of $C_l$.
The indicated averages are to be taken over the theoretical
ensemble of all possible anisotropy fields, of
which our observed CMB sky is but one realization.  
Since the harmonic coefficients are Gaussian variables
and the expansion is linear, it is clear that
the temperature values on the sky
are also Gaussian, and they therefore follow a multivariate 
Gaussian distribution (with an uncountably infinite 
number of variables, one for each position on the 
sky).  The covariance of temperatures separated
by an angle $\theta$ on the sky is given
by the {\em correlation function}
\begin{equation}
C(\theta) \equiv <\Delta(\hat{n}_1)\Delta(\hat{n}_2)>_{ens} = 
        \frac{1}{4\pi}\sum_{l} (2l+1) C_l P_l(\mu)
\end{equation}
where $P_l$ is the Legendre polynomial of order $l$
and $\mu = \cos\theta = \hat{n}_1\cdot\hat{n}_2$.  
The form of this equation, which follows
directly from Eq. (\ref{eq:defCl}),
is dictated by the statistical isotropy of the
perturbations -- the two--point correlation function can
only depend on separation.

        Observationally, one works with sky brightness 
integrated over the experimental beam
\begin{equation}
\Delta_b(\hat{n}_p) = \int\; d\Omega \Delta(\hat{n}) B(\hat{n}_p,\hat{n})
\end{equation}
where $B$ is the beam profile and $\hat{n}_p$ gives the
position of the beam axis.  The beam profile may or may not
be a sole function of $\hat{n}_p\cdot\hat{n}$, i.e., of
the separation between sky point and beam axis; if
it is, then this equation is a simple convolution on 
the sphere, and we may write
\begin{eqnarray}\label{eq:Cbeam}
C_{b}(\theta) \equiv <\Delta_b(\hat{n}_1)\Delta_b(\hat{n}_2)>_{ens}
        & = &\frac{1}{4\pi} \sum_l (2l+1) C_l \\
\nonumber
& & \hspace*{1cm} \times |B_l|^2 P_l(\mu)
\end{eqnarray}
for the beam--smeared correlation function, or covariance
between experimental beams separated by $\theta$.
The beam harmonic coefficients, $B_l$, are defined
by 
\begin{equation}
B(\theta') = \frac{1}{4\pi}\sum_l (2l+1) B_l P_l(\mu')
\end{equation}
with $\hat{n}_p\cdot\hat{n} = \cos\theta' = \mu'$.
For example, for a Gaussian beam, $B(\theta) = 1/(2\pi\sigma^2) 
e^{-\theta^2/2\sigma^2}$ and $B_l = e^{-l(l+1)\sigma^2/2}$.   

        Given these relations and a CMB map, it is now
straightforward to construct the likelihood function,
whose role is to relate the $\Npix$ 
observed sky temperatures, which we arrange in a 
{\em data vector} with elements $d_i \equiv 
\Delta_b(\hat{n}_i)$, to the model parameters,
represented by a {\em parameter vector}  $\vTeta$.
As advertised, for {\em Gaussian} fluctuations
(with Gaussian noise) this is simply a multivariate 
Gaussian:
\begin{equation}\label{eq:like1}
{\cal L}(\vTeta) \equiv {\rm Prob}(\vd|\vTeta)
        = \frac{1}{(2\pi)^{\Npix/2} |\mC|^{1/2}} e^{-\frac{1}{2}
        \vd^t \cdot \mC^{-1} \cdot \vd} 
\end{equation} 
The first equality reminds us that the likelihood function
is the probability of obtaining the data vector given the
model as defined by its set of parameters.  In this 
expression, $\mC$ is the pixel covariance matrix:
\begin{equation}\label{eq:covmat}
C_{ij} \equiv <d_id_j>_{ens} = T_{ij} + N_{ij}
\end{equation} 
where the expectation value is understood to be over the
theoretical ensemble of all possible universes realisable with
the same parameter vector.  The second equality separates
the model's pixel covariance, $\mT$, from the noise induced 
covariance, $\mN$.  According to Eq. (\ref{eq:Cbeam}),
$T_{ij} = C_b(\theta_{ij})$.  The parameters may 
be either the individual $C_l$ (or band--powers,
discussed below), or the fundamental cosmological
constants, $\Omega, H_o$, etc...  In the former
case, Eq. (\ref{eq:Cbeam}) shows how the parameters
enter the likelihood; in the latter situation,
the parameter dependence enters through detailed
relations of the kind $C_l[\vTeta]$, specified
by the adopted model (e.g., Inflation).  Notice
that if one only desires to determine the $C_l$, then
only the assumption of Gaussianity is required.

        Many experiments report temperature 
{\em differences}; and even if the starting
point is a true map, one may wish to subject it
to a linear transformation in order to define
bands in $l$--space over which power estimates
are to be given.  Thus, it is useful to generalize
our approach to arbitrary homogeneous, linear
data combinations, represented by a transformation
matrix $\mA$: $\vd^\prime = \mA\cdot \vd$.
Since the transformation is linear, the
new data vector retains a multivariate
Gaussian distribution (with zero mean), but 
with a modified covariance matrix:
$\mC^\prime = \mA \cdot \mC \cdot \mA^t$.
As a consequence, the transformed pixels, 
$\vd'$, may be treated
in the same manner as the originals, and
so we will hereafter use the term {\em generalized
pixels} to refer to the elements of a general
data vector which may be either real sky pixels or
some transformed version thereof.  
The elements of the new theory covariance matrix
are (using the summation convention)
\begin{equation}\label{eq:defW}
T^\prime_{ij} = A_{im}A_{jn}T_{mn} = \frac{1}{4\pi}
        \sum_l (2l+1) C_l W_{ij}(l)
\end{equation}
where $W_{ij}(l) \equiv A_{im}A_{jn}P_l(\mu_{mn})|B_l|^2$.
The {\em window function} is usually 
defined as $W_{ii}(l)$, i.e., the 
diagonal elements of a more general matrix
$\mW(l)$.  Normally, one tries to find
a transformation which leads to a strongly
diagonal $\mW(l)$ {\em and} diagonal noise matrix
(see comment below).   

        An example is helpful.  Consider a 
simple, single difference $\Delta_{diff}
\equiv \Delta_b(\hat{n}_1)-\Delta_b(\hat{n}_2)$, 
whose {\em variance} is given by 
$<\Delta_{diff}^2>_{ens} = 2[C_b(0) - C_b(\theta)]$.
This may be written in terms of multipoles as
\begin{equation}
<\Delta_{diff}^2> = \frac{1}{4\pi} \sum_l (2l+1) C_l  
        \left\{ 2 |B_l|^2 \left[1 - P_l(\mu)\right] \right\}
\end{equation}
identifying the diagonal elements of $\mW$ as
the expression in curly brackets.  Notice that 
the power in this variance is localized
in $l$--space, being bounded towards large $l$ by 
the beam smearing and towards small $l$ by the difference. 
The off--diagonal
elements of $\mC$ depend on the relative positions
and orientations of the differences on the sky;
in general these elements are not expressible as
simple Legendre series.

        Band--powers are defined via Eq. (\ref{eq:defW}).
One reduces the set of $C_l$ contained within the
window to a single number by adopting a spectral form.
The so--called {\em flat band--power}, $\fbP$, is established by
using $C_l \equiv 2\pi (\fbP)^2/[l(l+1)]$, leading to 
\begin{equation}\label{eq:T}
\mT = \frac{1}{2} \fbP^2 \sum_l \frac{(2l+1)}{l(l+1)} \mW(l)
\end{equation}
In this fashion, we may write Eq. (\ref{eq:like1}) in 
terms of the band--power and treat the latter as
a parameter to be estimated.  This then becomes the 
band--power likelihood function, ${\cal L}(\fbP)$.  
One obtains the points shown in Figure 1 by maximizing this 
likelihood function; the errors are typically found by in a 
Bayesian manner, by integration over ${\cal L}$ with a 
uniform prior.  Notice that the variance due to the finite
sample size (i.e., the sample variance, but also known as 
cosmic variance when one has full sky coverage)  
is fully incorporated into the analysis --
the likelihood function ``knows'' how many pixels
there are.

        An important remark at this stage concerns the
construction of Figure 1.  We see here that this
figure is only valid for Gaussian perturbations, because
it relies on Eq. (\ref{eq:like1}), which assumes 
Gaussianity at the outset.  If the sky fluctuations
are non--Gaussian, then these estimates must all 
be re--evaluated based on the true nature of the sky 
fluctuations, i.e., the likelihood function in Eq. (\ref{eq:like1})
must be redefined.  The same comment applies to any
experiment which has an important non--Gaussian noise 
component -- the likelihood function must incorporate
this aspect in order to properly yield the power estimate
and associated error bars.  

        What is the {\em raison d'\^etre} for
these band powers?  The likelihood function is clearly
greatly simplified if we can find a 
transformation $\mA$ which diagonalizes
$\mC$ (signal plus noise).  This can be done
for a given model, but because $\mC$ 
depends on the model parameters, there is in
general no unique such transformation valid
for all parameter values. 
The one exception is for an ideal experiment 
(no noise, or uniform, uncorrelated noise) with 
full--sky coverage -- in this case
the spherical harmonic transformation 
is guaranteed, by Eq. (\ref{eq:defCl}), 
to diagonalize $\mC$ for any and all values of the model 
parameters.  This linear transformation is
represented by a matrix $A_{ij} \equiv
Y_i(\hat{n}_j)$, where $i= l^2 + l + m + 1$ is 
a unidimensional index for the pair $(l,m)$.  
It is the role of band--powers to approximately
diagonalize the covariance matrix in
more realistic situations, where sky
coverage is always limited and noise is 
never uniform (and sometimes correlated),
and in such a way as to concentrate the
power estimates in as narrow bands as
possible.  Since this is not possible for 
arbitrary parameter values,  
in practice one adopts a fiducial model
(particular values for the parameters)
to define a transformation $\mA$
which compromises between the desires
for narrow and independent bands 
(Bond 1995, Tegmark et al. 1997, Tegmark 1997, 
Bunn \& White 1997).

\subsection{Motivating an Ansatz}
        
        Given a set of band--powers, how should
one proceed to constrain the fundamental cosmological
parameters, denoted in this subsection by $\vTeta$?
If we had an expression for ${\cal L}(\vfbP)$,
for our set of band--powers $\vfbP$, then
we could write ${\cal L}(\vfbP) = Prob(\vd|\vfbP) 
= Prob(\vd|\vfbP[\vTeta]) = {\cal L}(\vTeta)$.
Thus, our problem is reduced to finding an expression for 
${\cal L}(\vfbP)$, but as we have seen, this is a complicated 
function of $\vfbP$, requiring use of all the measured pixel 
values and the full covariance matrix with noise --
the very thing we are trying to avoid.
Our task then is to find an approximation for
${\cal L}(\vfbP)$.   
In order to better understand the general form
expected for ${\cal L}(\vfbP)$, we shall proceed by
first considering a simple situation in which we may 
find an exact analytic expression for this function.  
We are guided by the observation that the covariance
matrix may always be diagonalized around an adopted
fiducial model.  Although this remains strictly 
applicable only for this model, we imagine that
the likelihood function could be approximated as 
a simple product of one--dimensional Gaussians 
near this point in parameter space.  If we further
suppose that the diagonal elements of the covariance
matrix (its eigenvalues) are all identical, we can find
a very manageable analytic expression for the
likelihood in terms of the best power estimate.
We will then pose this general form as an ansatz for
more realistic situations, one which we shall test in the
following section.  We return to these remarks after
developing the ansatz. 

        Consider, then, a situation in which 
the band temperatures (that is, generalized pixels which
are the elements of the general 
data vector $\vd'$) are independent random variables 
($\mC$ is diagonal) and that the experimental noise
is spatially uncorrelated and uniform:
\begin{eqnarray}\label{eq:simplecov}
C_{ij} = (\sigM^2 + \sigN^2)\delta_{ij}
\end{eqnarray}
where $\sigM^2$ is the model--predicted variance and
$\sigN^2$ is the constant noise variance.
For simplicity, we assume that all diagonal
elements of $\mW$ are the same, implying that
$\sigM^2$ is a constant, independent of $i$.
We discuss shortly the nature of such
a data vector in actual observational set--ups.
This situation is identical to one where
$\Npix$ values are randomly selected from
a single parent distribution described by
a Gaussian of zero mean and variance 
$\sigM^2 + \sigN^2$.  The band--power we wish 
to estimate is proportional
to the model--predicted variance according to
(i.e., Eq. \ref{eq:T})
\begin{equation}\label{eq:Rband}
\sigM^2 = \fbP^2 \times \frac{1}{2} \sum_l 
        \frac{(2l+1)}{l(l+1)} W_{ii}(l) \equiv 
        \fbP^2 {\cal R}_{band}
\end{equation}
(independent of $i$),
and we know that in this situation the maximum 
likelihood {\em estimator} for the model--predicted 
variance is simply
\begin{equation}\label{eq:EsigM}
[\EsigM]^2 = \frac{1}{\Npix} \sum_{i=1}^{\Npix} d_i^2 - \sigN^2
        \equiv [\EfbP]^2 {\cal R}_{band}
\end{equation}
as follows from maximizing the likelihood function
\begin{displaymath}
{\cal L}(\sigM)  =  \frac{1}{[2\pi(\sigM^2+\sigN^2)]^{\Npix/2}}
        e^{-\frac{\Npix(\EsigM^2+\sigN^2)}{2(\sigM^2+\sigN^2)}}
\end{displaymath}
Notice that this is {\em a function of $\sigM$}, which
peaks at the best estimate $\EsigM$, and whose form
is specified by the parameters $\EsigM$, $\sigN$ 
and $\Npix$.  To obtain the likelihood function for
the band--power, we simply treat this as a function of 
$\fbP$, using Eq. (\ref{eq:Rband}), parameterized by
$\EfbP$, $\sigN$ and $\Npix$:
\begin{eqnarray}\label{eq:likeTfb}
{\cal L}(\fbP)  & = & \frac{1}{[2\pi(\fbP^2{\cal R}_{band}+\sigN^2)]
        ^{\Npix/2}} \\
\nonumber 
& & \hspace*{2cm} \times e^{-\frac{\Npix(\EfbP^2{\cal R}_{band}+\sigN^2)}
        {2(\fbP^2{\cal R}_{band}+\sigN^2)}}\\ 
\nonumber
\\ 
\nonumber
& \equiv & G(\fbP;\EfbP,\sigN,\Npix)
\end{eqnarray}
It clearly peaks at $\EfbP$.  Thus, in this ideal case, 
we have a simple band--power likelihood function, with corresponding
best estimator, $\EfbP$, given by Eq. (\ref{eq:EsigM}).

        Although not immediately relevant to our present
goals, it is all the same instructive to consider the {\em
distribution} of $\EfbP$. 
This is most easily done by noting that the quantity
\begin{equation}
\chi^2_{\Npix} \equiv  
        \sum_{i=1}^{\Npix} \frac{d_i^2}{\sigM^2 
        + \sigN^2}
\end{equation}
is $\chi^2$--distributed with $\Npix$ degrees of freedom.
We may express the maximum likelihood estimator
for the band--power in terms of this quantity
as
\begin{equation}\label{eq:trans}
\EfbP^2 =  {\cal R}_{band}^{-1} 
        \left[\frac{(\sigM^2 + \sigN^2)}{\Npix} \chi^2_{\Npix}-\sigN^2\right]
\end{equation}
From $<\chi^2_{\Npix}>=\Npix$,
we see immediately that the estimator is unbiased 
\begin{displaymath}
<\EfbP^2>_{ens} = {\cal R}_{band}^{-1}\sigM^2 = \fbP^2
\end{displaymath}
Its variance is 
\begin{eqnarray*}
\nonumber
Var(\EfbP^2) & = & {\cal R}_{band}^{-2}
        \frac{(\sigM^2 + \sigN^2)^2}{\Npix^2} Var(\chi^2_{\Npix}) \\
\nonumber
\\
\nonumber
& = & 2 {\cal R}_{band}^{-2} (\sigM^2 + \sigN^2)^2/\Npix
\end{eqnarray*}
explicitly demonstrating the influence of sample/cosmic variance
(related to $\Npix$).

        All the above relations are {\em exact} 
for the adopted situation -- Eq. (\ref{eq:likeTfb}) 
is the {\em complete} likelihood function for the band--power
defined by the {\em generalized} pixels satisfying
Eq. (\ref{eq:simplecov}).  Such a situation could
be practically realized on the sky by observing well 
separated generalized pixels to the same noise level;
for example, a set of double differences scattered
about the sky, all with the same signal--to--noise.  
This is rarely the case, however, as scanning strategies
must be concentrated within a relatively small 
area of sky (one makes maps!).  This creates important
off--diagonal elements in the theory covariance
matrix $\mT$, representing correlations between
nearby pixels due to long wavelength perturbation
modes.  In addition, the noise
level is quite often not uniform and sometimes 
even correlated, adding off--diagonal elements
to the noise covariance matrix. Thus, the simple
form proposed in Eq. (\ref{eq:simplecov}) is
never achieved in actual observations. 
Nevertheless, as mentioned, even in this case one
could adopt a fiducial theoretical model
and find a transformation
$\mA$ which diagonalizes the full covariance 
matrix $\mC$, thereby regaining one important 
simplifying property of the above ideal situation.
The diagonal elements
of the matrix are then its eigenvalues.
Because of the correlations in the original
matrix, we expect there to be fewer significant
eigenvalues than generalized pixels;  this 
will be relevant shortly.  One could then
work with a reduced matrix consisting
of only the significant eigenvalues, an
approach reminiscent of the signal--to--noise
eigenmodes proposed by Bond (1995), and 
also known as the Karhunen-Loeve transform
(Bunn \& White 1997, Tegmark et al. 1997).
There remain two technical difficulties:
the covariance matrix does not remain
diagonal as we move away from the adopted fiducial
model by varying $\fbP$ -- only when this band--power
corresponds to the fiducial model is the
matrix really diagonal.  The second 
complicating factor is that the eigenvalues
are not identical, which greatly simplified
the previous calculation.  

        All of this motivates us to examine the possibility 
that a likelihood function of 
the form (\ref{eq:likeTfb}) could be applied,
with appropriate redefinitions of $\Npix$ and
$\sigN$.  We therefore proceed by renaming
these latter $\nu$ and $\beta$, respectively,
and treating them as parameters to be adjusted
to best fit the full likelihood function.  
Thus, given an actual band--power estimate, $\fbP^{(o)}$ 
(i.e., an experimental result), {\em we propose 
$G(\fbP;\fbP^{(o)},\beta,\nu)$
as an ansatz for the band--power likelihood function,
with parameters $\nu$ and $\beta$}:\\
\begin{eqnarray}\label{eq:approx}
{\cal L}(\fbP) & \propto & X^{\nu/2} e^{-X/2} \\
\nonumber
\\
\nonumber
 X[\fbP] & \equiv & \frac{([\fbP^{(o)}]^2 + \beta^2)}{([\fbP]^2 +
        \beta^2)}\nu \\
\nonumber
\end{eqnarray}
We have only two parameters -- $\nu$ and $\beta$ -- 
to determine in order
to apply the ansatz.  This can be done if two confidence
intervals of the complete likelihood function are known in
advance.  For example, suppose we were given both the 
68\% ($\sigma^+_{68}$ \& $\sigma^-_{68}$) and 95\% 
($\sigma^+_{95}$ \& $\sigma^-_{95}$)
confidence intervals; then we could fix
the two parameters with the equations
\begin{eqnarray}\label{eq:sixty8}
0.68 & = & \frac{\int_{\fbP^{(o)}-\sigma^-_{68}}^{\fbP^{(o)}+\sigma^+_{68}}
        d[\fbP]\; {\cal L}(\fbP)}
        {\int_0^\infty d[\fbP]\; {\cal L}(\fbP)} \\
\nonumber
\\
\label{eq:ninty5}
0.95 & = & \frac{\int_{\fbP^{(o)}-\sigma^-_{95}}^{\fbP^{(o)}+\sigma^+_{95}}
        d[\fbP]\; {\cal L}(\fbP)}
        {\int_0^\infty d[\fbP]\; {\cal L}(\fbP)} \\
\nonumber
\end{eqnarray}
We shall see in the next section (Figures 2--7) that this 
produces excellent approximations.  This is the main result of
this {\em paper}.

\begin{figure*}\label{fig_saskcomp1}
\begin{center}
\resizebox{\hsize}{!}{\includegraphics[totalheight=8cm,
        width=16cm]{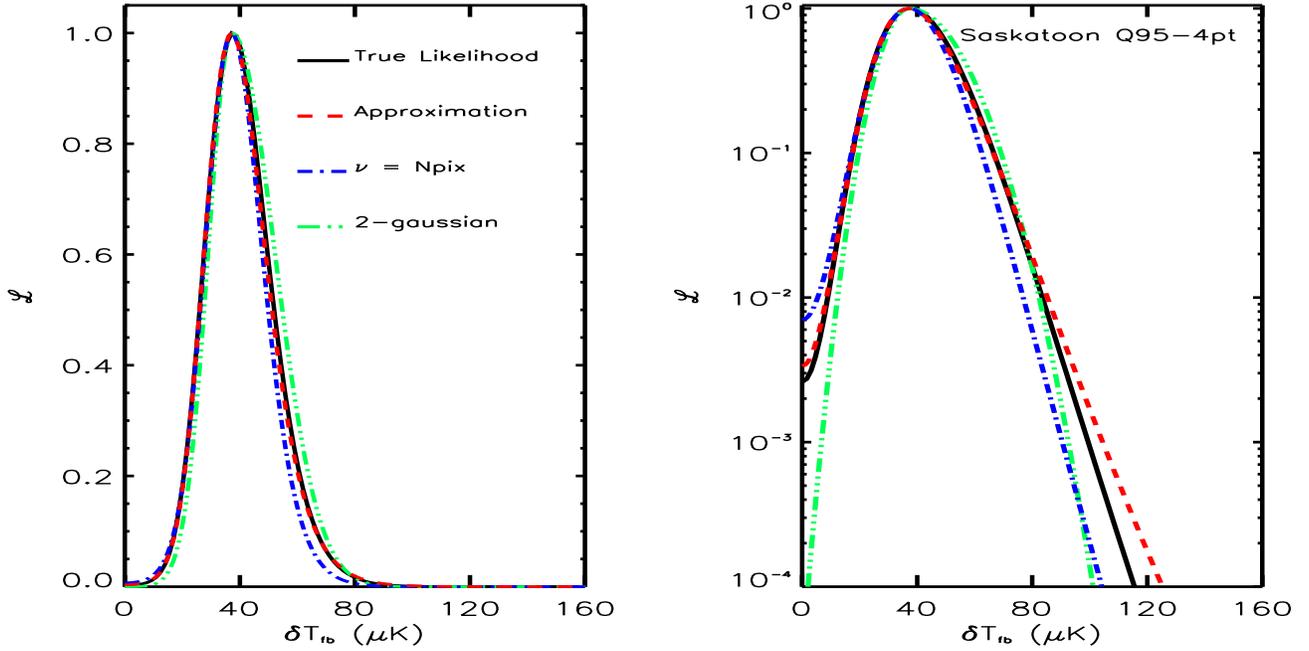}}
\end{center}
\caption{Comparison to the Saskatoon Q--band
1995 4--point difference.  The value of the
likelihood is plotted as a function of the
band--power, $\fbP$, in both linear (left) and logrithmic
(right) scales.  The solid (black) curve in each case gives the
true likelihood function, while the dashed (red) curve
corresponds to the proposed approximation based on 
two confidence intervals.  The dot--dashed (blue) curve
is the ansatz with $\nu=\Npix=24$ and $\beta$ adjusted
to the 68\% confidence interval (see text).
A ``2--winged Gaussian'' with  different positive--going and 
negative-going errors is shown as the three--dotted--dashed (green) 
curve.  All curves have been normalized to unity at their peaks.} 
\end{figure*}

\begin{figure*}\label{fig_saskcomp2}
\begin{center}
\resizebox{\hsize}{!}{\includegraphics[totalheight=8cm,
        width=16cm]{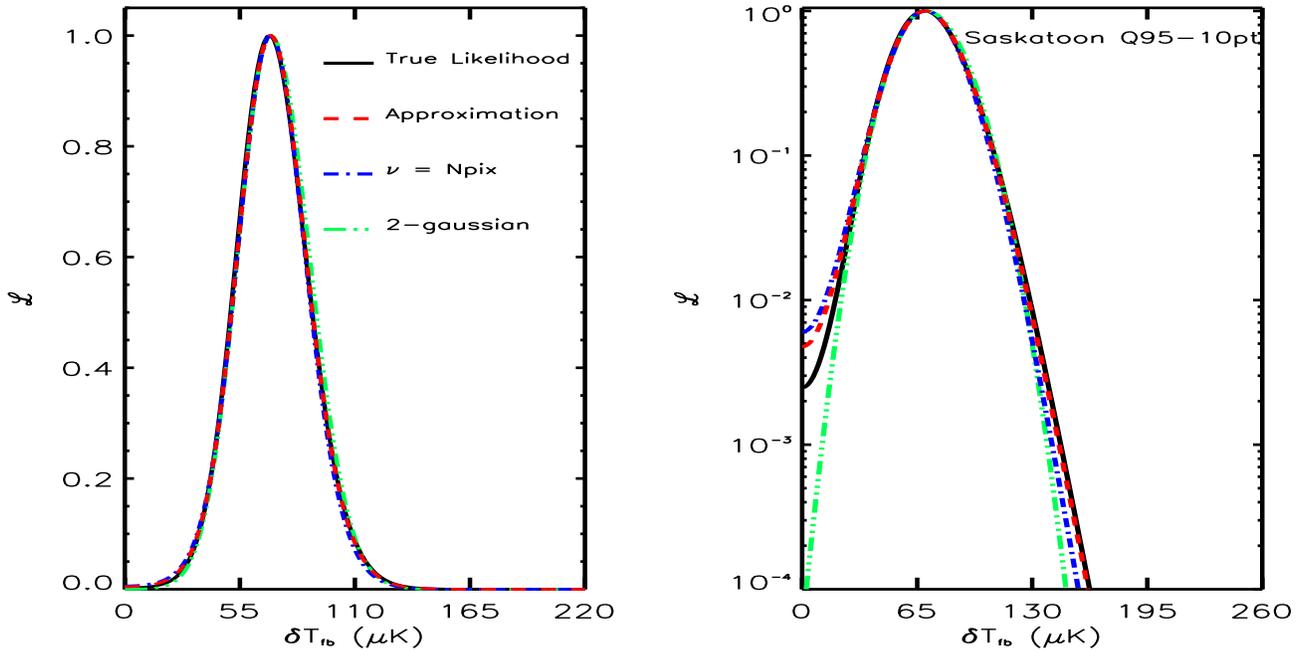}}
\end{center}
\caption{Comparison to the Saskatoon Q--band
1995 10--point difference.  The line-styles are
the same as in the previous figure; here 
$\Npix=48$ for the dot--dashed (blue) line.} 
\end{figure*}

        Unfortunately, most of the time only the 68\%
confidence interval is reported along with an
experimental result (we hope that in the future 
authors will in fact supply at least two confidence
intervals).  Is there any way to proceed in this
case?  For example, one could try to judiciously
choose $\nu$ and then adjust $\beta$ with Eq. (\ref{eq:sixty8}).
The most obvious choice for $\nu$ would be $\nu=\Npix$, 
although from our previous discussion, we expect this
to be an upper limit to the number of significant 
degrees--of--freedom (the significant eigenvalues of 
$\mC$), due to correlations between pixels.
The comparisons we are about to make in the following
section show that a smaller number of effective pixels (i.e.,
value for $\nu$) is in fact required for a good fit to
the true likelihood function.  One could try other games,
such as setting $\nu\equiv$ (scan length)/(beam FWHM) 
for unidimensional scans.  This also seems reasonable, and
certainly this number is less than or equal to the actual
number of pixels in the data set, but we have found
that this does not always work satisfactorily.  The availability of a
second confidence interval permits both parameters,
$\nu$ and $\beta$, to be unambiguously determined
and in such a way as to provide the best possible 
approximation with the proposed ansatz.  

        Bond et al. (1998) have recently examined the nature
of the likelihood function and discussed two 
possible approximations.  The form of the ansatz just presented is
in fact identical to one of their proposed approximations,
parameterized by $x$ and $G$.  These parameters 
are simply related to our $\nu$ and $\beta$ as follows:
$x=\beta^2$ and $G=\nu$. 

        Notice that the above development and motivation
for the ansatz essentially follow for a single band--power.
A set of uncorrelated power estimates is then easily treated
by simple multiplication.  However, the approximation as
proposed does not simultaneously account for several
{\em correlated} band--powers, and it's accuracy is
therefore limited by the extent to which such inter--band
correlations are important in a given data set.
As a further remark along these lines,
we have noted that flat--band estimates of any kind,
be it from a complete likelihood analysis or not, do
not always contain all relevant experimental information,
(Douspis et al. 2000); any method based on their use
is then fundamentally limited by nature of the lost
information.                      

        The only way to test the ansatz is, of course,
by direct comparison to the full likelihood function 
calculated for a number of experiments.  If it appears
to work for a few such cases, then we may hope
that it's general application is justified.
We now turn to this issue.

\section{Testing the approximation}

        In order to quantitatively test 
the proposed ansatz, we have 
calculated the complete likelihood function for 
several experiments.  Our aim will be
to compare the true likelihoods to 
the approximation.  Figures 2--5 summarize our comparisons
with the Saskatoon and MAX data sets.  For the 
Saskatoon and MAX experiments, we compare the
approximation directly 
to the band--power likelihood functions. 
In all cases, the complete likelihood functions
have been calculated as outlined in Section
2 above.   

\begin{figure*}\label{fig_maxcomp}
\begin{center}
\resizebox{\hsize}{!}{\includegraphics[totalheight=8cm,
        width=16cm]{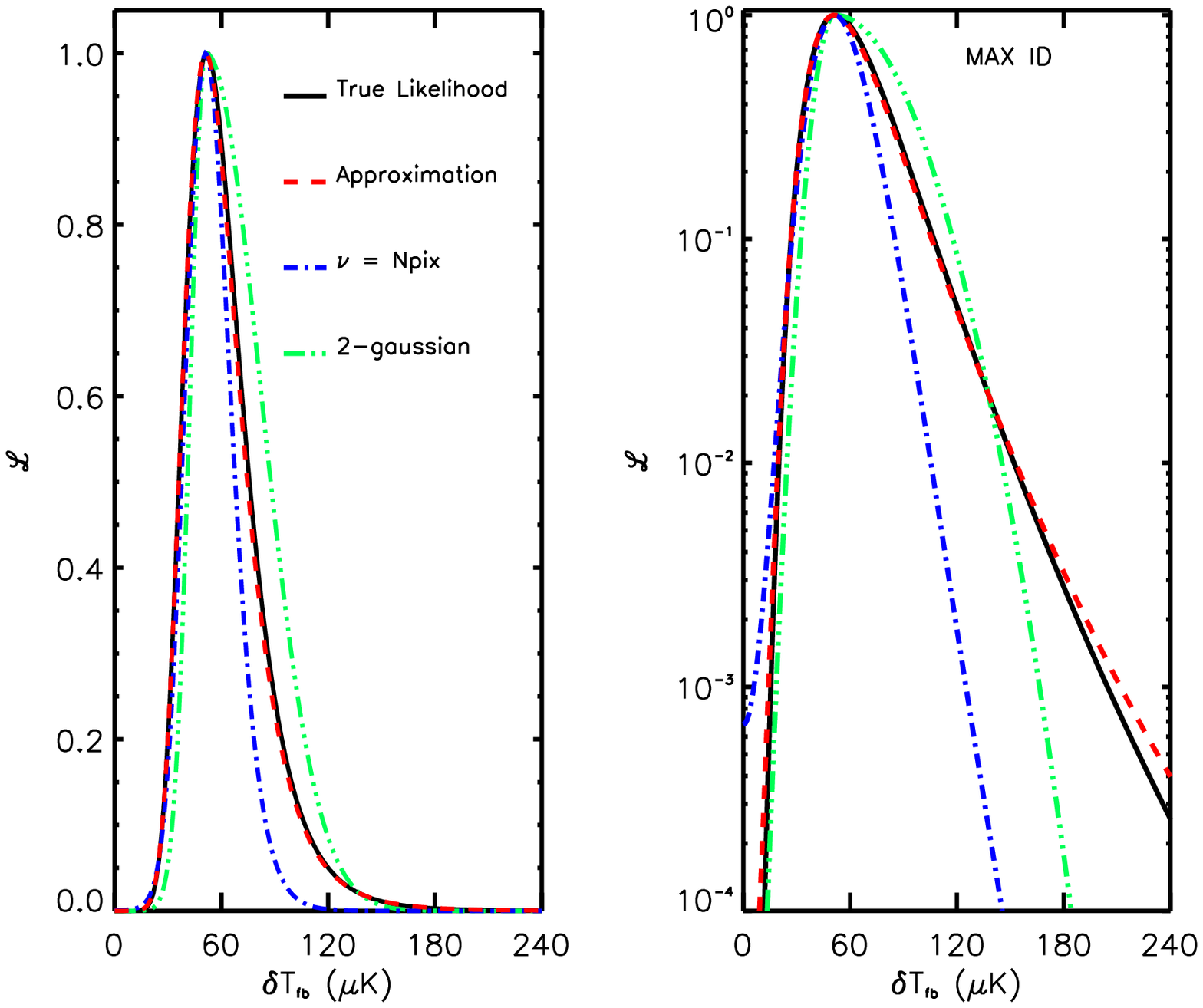}}
\end{center}
\caption{Comparison to the MAX ID likelihood function.
This is the combined likelihood for the 3 frequency channels,
3.5, 6 and 9 $cm^{-1}$.  Linestyles are the same as in the previous
two figures, and here $\Npix=21$ applies to the dot--dashed (blue) line.}
\end{figure*}

        The first comparison will be made to 
the Saskatoon Q--band 1995 4--point and 10--point
differences (experimental information can be
found in Netterfield et al. 1997;
all relevant information concerning the
experiment can be found on the group's web 
page\footnote{{\tt http://pupgg.princeton.edu/$\sim$cmb/skintro/\\
sask\_intro.html}} ;
for useful and detailed information on a number of experiments, see Caltech's
web page\footnote{{\tt http://crunch.ipac.caltech.edu:8080/imbarc/}}).
This particular choice of window functions was arbitrary.
The approximation, applied using the constraints
(\ref{eq:sixty8}) and (\ref{eq:ninty5}), is shown in Figures 2 and 3 
as the dashed (red) curve.  We see that it provides    
a good representation of the complete likelihood functions,
traced by the solid (black) curves in each figure; 
in fact, the fit is truly spectacular for 
the 10--point difference.
Taking as a benchmark the rule--of--thumb that
1, 2 and 3 $\sigma$ confidence intervals may be
estimated by $2\Delta\ln{\cal L} = 1$, 4 and 9,
respectively, we see that the approximation 
reproduces almost perfectly all of these, and
more.  

        Consider now setting $\nu=\Npix=24$ and $48$, for
the 4--point and 10--point differences, respectively,
and then adjusting
$\beta$ to the 68\% confidence interval.  In so doing, we obtain
the dot--dashed (blue) curves, which in fact are not too
bad in both cases.  These values of $\Npix$ should
be compared to the values of $\nu=16$ and $41$ found previously
by adjusting to two confidence intervals.  
Thus, we see that the effective 
number of degrees--of--freedom describing these Saskatoon 
likelihood functions is indeed $\nu\le \Npix$, as we expected from 
the above discussion.  

        Finally, the 3--dot--dashed (green) 
curves show ``2--winged''
Gaussians with separate positive-- and negative--
going variances, sometimes employed in
traditional $\chi^2$--analyses.  This 
is also a fare representation of the
two likelihood functions, although the proposed
ansatz does perform slightly better.  We will return to this
point, but we should not be too surprised that
the Gaussian works reasonably well when, as
here, $\nu$ becomes large (all the same, notice
that the curves are not symmetric and that
a single Gaussian, with a single $\sigma$,
would not fare particularly well).

        Comparison to the MAX experiment is 
shown in Figure 4 for the region ID (experimental details
can be found in Tanaka et al. 1996); we have combined
all three frequency channels to construct the complete likelihood
function.  The scan strategy consisted in taking $\Npix=21$ single 
differences aligned along a unidimensional scan.
Once again, the approximation, applied using Eqs. (\ref{eq:sixty8}) and
(\ref{eq:ninty5}), supplies
an excellent representation of the likelihood function,
down to values well below ``$3\sigma$'' (0.01 of the peak).  
The effective number of degrees--of--freedom is 
$\nu=8.5$, demonstrating
again that $\nu\le\Npix$.  Here, the
difference is rather large, due to the significant overlap
between adjacent pixels along the scan, and we see that
the ansatz with $\nu=\Npix$ does not produce a good approximation.

\begin{figure*}\label{figsaskbin3}
\begin{center}
\resizebox{\hsize}{!}{\includegraphics[totalheight=8cm,
        width=15cm]{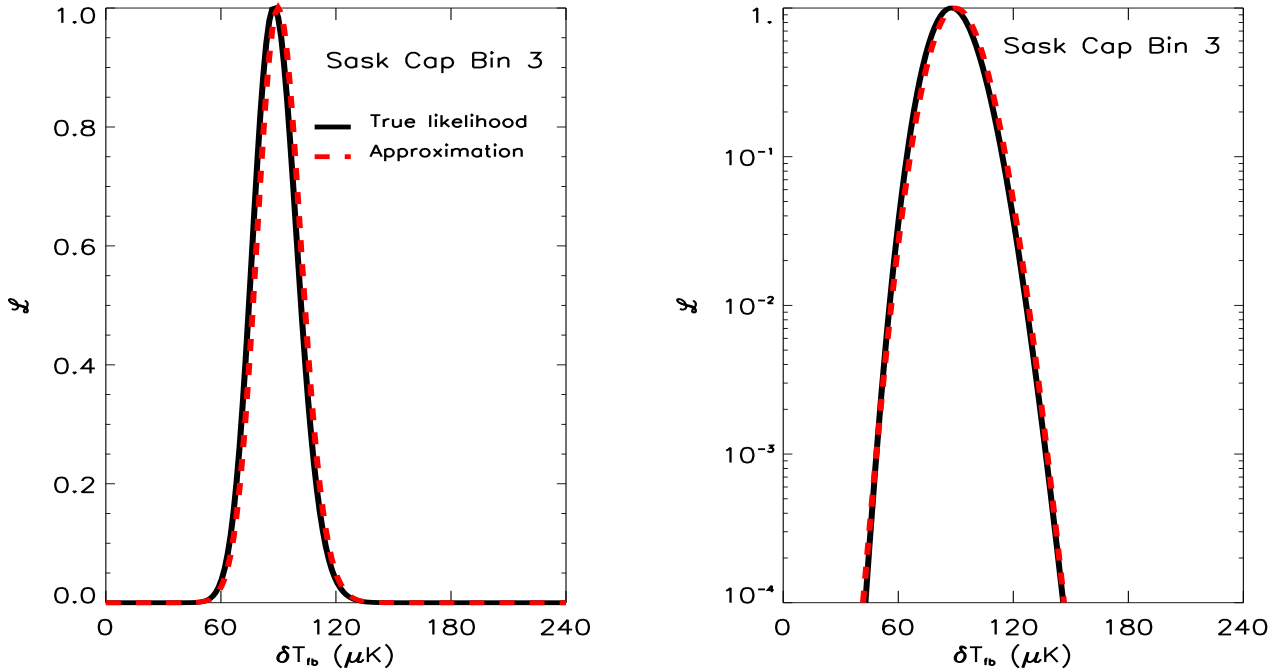}}
\end{center}
\caption{Comparison to a Saskatoon bin consisting of the 10--,
11-- and 12--point differences (bands) from the K--band 1994,
Q--band 1994 and Q-band 1995 CAP data.  As previously, the solid
black line shows the true likelihood function, while the
dashed (red) curve displays the approximation based on two
confidence intervals.  This Figure demonstrates that
the approximation works well even when several individual bands
are combined to form a band--power likelihood.}
\end{figure*}

        Could there a way to proceed if only one confidence 
interval is given?  This would require a choice for
one of the parameters, say $\nu$, based on some knowledge of the
scan strategy.  We have just seen that for MAX $\nu=\Npix$
leads to a bad representation of the likelihood function.  
One might be tempted to try instead $\nu=$
(scan length)/(beam FWHM)$=8.8$ , which is in fact very 
close to the best value of $\nu$ found from adjusting
to two confidence intervals.  Although this is successful
in this case, it is nevertheless
guess--work, the problem being that
it is really not clear if there is a unique rule for 
judiciously choosing $\nu$.  For Saskatoon,
$\nu=\Npix$ worked reasonably well, while here it
does not, something much less being required because
of the significant redundancy in the scan.  We have found
that it is difficult to justify a priori 
a general rule for choosing $\nu$ when lacking two
confidence intervals.  The most sure way of finding
the effective number of degrees--of--freedom to be used
in the ansatz remains the use of two confidence intervals, \
via Eqs. (\ref{eq:sixty8}) and (\ref{eq:ninty5}). 

        A noteworthy aspect of this MAX likelihood function
is its asymmetry, i.e., it is manifestly non--Gaussian.  
Even a ``2--winged'' Gaussian is clearly a very bad 
representation.  As  
the number of statistically independent elements 
entering the power estimation increases, we should
expect the likelihood function to approach a Gaussian
distribution.  The question is, what is meant by
{\em statistically independent elements}?  It is 
obviously {\bf not} something like $2l+1$, for MAX
covers multipoles near 100; rather, as we have
argued above, it is really the parameter $\nu$
which measures this, what we have been calling
the effective number of degrees--of--freedom.  The
fact that $\nu \le \Npix$ tells
us that the number of generalized pixels is
an {\em upper limit} to this number degrees--of--freedom
determining the non--Gaussian nature of the likelihood
function.  We make the connection to the familiar $2l+1$--rule
only when we have full--sky coverage and bands consisting
of single multipoles; then, the number of generalized
pixels defining each (single multipole) band corresponds
to $2l+1$.  In the general case, 
it is more useful and correct to reason with the number of pixels
(really, $\nu$).
We may also conclude from this that although experiments
with relatively large sky coverage should provide 
Gaussian likelihood functions on scales much smaller than
the survey area, band--power estimates on scales
approaching the survey area will always be non--Gaussian.
The proposed ansatz represents a substantial improvement
over either a single or ``2--winged'' Gaussian in such
cases.

        These comparisons focus on simple
cases where the power over a single band defined
by the observing strategy is to be estimated,
although in the MAX case the analysis did
include three frequency channels
simultaneously.  A more subtle test of
the approximation is its extension
to a power estimate over several bands
defined by {\em different} window functions.
Such is the situation presented by the five 
standard Saskatoon power bins.  Each {\em bin}
comprises several {\em bands}, of the type
considered above, and the bin power is 
estimated using the joint likelihood 
of the contributing bands, including all
band--band correlations.  One could worry
that the information carried by several
bands might not be adequately incorporated
by the two parameters of the ansatz.

     In Figure 5 we compare the approximation
to the likelihood function of a combination of
10--, 11-- and 12--point differences.  Included
are the K--band 1994 and Q--band 1994 and 1995 
CAP data.  The true likelihood function for this 
bin is calculated from the complete covariance matrix
accounting for all correlations, and the approximation
was fit using two confidence intervals. 
Even in this more complicated situation
we see that the ansatz continues to work quite well,
once the appropriate best power estimate and
errors for the complete bin are used to 
find $\nu$ and $\beta$.

        It is on the basis of such comparisons
that we believe the proposed ansatz and
method of application produces acceptable
likelihood functions.  Besides the comparisons shown
here, we have also tested the approximation against
11 other complete likelihood functions, all kindly
provided by K. Ganga; these comparisons may be viewed on our 
web page\footnote{{\tt http://astro.u-strasbg.fr/Obs/COSMO/CMB/}}.
The approximation works well in all cases.
We emphasize again that 
the particular value of the proposed ansatz resides in its 
simplicity -- we obtain
very good approximations with little effort.

\section{Conclusion}

        Study of CMB temperature fluctuations have over the
short interval of time since their discovery become the cosmological
tool with the greatest potential for determining the values
of the fundamental cosmological constants.  The present
data set is already capable of eliminating some regions of
parameter space, and this is only a fore-taste of what
is to come.  Experimental results are often quoted as
band--power estimates, and for {\em Gaussian} sky fluctuations,
these represent a complete description of an observation.
Because there are far fewer band--powers than pixel values
for any given experiment, the reduction to band--powers has
been called ``radical compression'' (Bond et al. 1998); and
as the number of pixels explodes with the next instrument
generations, this kind of compression will become increasingly 
important in any systematic analysis of parameter constraints.

        For these reasons, it is extremely useful to develop 
statistical methods which take as their input power estimates.   
Since most standard methods use as a starting point the
likelihood function, one would like to have a simple
expression for this quantity given a power estimate -- one
that does not require manipulation of the entire observational
pixel set.  One difficulty is that even for Gaussian sky
fluctuations, the band--power likelihood function
is not Gaussian, most fundamentally because
the power represents an estimate of the {\em variance} of the
pixel values.  For any fiducial model,
the data covariance matrix can be diagonalized and the
likelihood function near this point in parameter
space expressed as a product of individual Gaussians
in the data elements
(this is strictly speaking only possible for the model
in question).  This consideration 
lead us to examine the ideal situation where the
eigenvalues of $\mC$ were all identical, for
which we can analytically find the exact form of the
likelihood function in terms of the best power estimate.
Using this as motivation, we have proposed the 
same functional form for band--power likelihood functions,
Eq. (\ref{eq:approx}), as an ansatz in more general cases. 
It contains two free parameters, 
$\nu$ and $\beta$, which may be uniquely determined if two
confidence intervals of the full likelihood function (the
thing one is trying to fit) are known; for example,
the 68\% and 95\% confidence intervals (Eqs. \ref{eq:sixty8}
and \ref{eq:ninty5}).  We have 
seen that the resulting approximate distributions
match remarkably well the complete likelihood functions
for a number of experiments -- those discussed here as well
as 11 others (calculated by K. Ganga and B. Ratra).  All of these 
comparisons may be viewed at our web 
site\footnote{{\tt http://astro.u-strasbg.fr/Obs/COSMO/CMB}}, where
we also plan to provide and continually up--date the
appropriate parameter values $\nu$ and $\beta$ for
each published experiment.

        Although at least one confidence interval is 
normally given in the literature (usually at 68\%), a second 
confidence interval is rarely quoted.  To aid the kind
of approach proposed here, we would ask that in the future
experimental band--power estimates be given with at least
two likelihood--based confidence intervals (additional
intervals, such as 99.8\%, would allow one to fit
other functional forms with 3 free parameters). 
This remains the surest way of finding the effective
number of degrees--of--freedom of the likelihood, $\nu$.
An otherwise a priori choice for this number appears difficult, 
among other things because it depends on the nature
of the scan strategy.  We have noted in this light
that $\nu\le \Npix$, precisely because of correlations between
pixels, which depend on the scan geometry.

        One important aspect of the approximate nature of
the proposed method is its inability to account for
correlations between several band--powers.  When
analyzing a set of band--powers, one is
obliged to simply multiply together their respective
approximate likelihood functions.  The accuracy of the approximation
is thus limited by the extent to which inter--band
correlations are important.  Although one's desire is
to give experimental results as independent power
estimates, this is not always possible.
Furthermore, and as discussed in
Douspis et al. (2000), the very use of flat--band powers
may lead to a loss of relevant experimental information
otherwise contained in the original pixel data.  The
accuracy of any method based on their use is thus
additionally limited by the importance of this
lost information.  These limitations define in
practice the approximate nature of the proposed method.

        Another important point to make is that the
approximation is extremely easy to use, as easy as 
the (inappropriate) $\chi^2$ method; and for 
experiments with a small number of significant 
degrees--of--freedom, it represents
a substantial improvement over the latter.   This
is the case, for example, with the MAX ID likelihood 
function, and it will always be the case when
estimating power on the largest scales of
a survey.  When the effective number of 
degrees--of--freedom becomes large, a Gaussian
becomes an acceptable approximation, and the
gain in using the proposed ansatz is less 
significant.  Nevertheless, the approximation's
facile applicability promotes its use even in
these cases.  In the future, we will 
apply the proposed approximation in a systematic study
of parameter constraints and for a test of the Gaussianity
of the CMB fluctuations.

\begin{acknowledgements}
        We are very grateful to K. Ganga and B. Ratra for so kindly 
providing us with an
additional 11 likelihood functions with which to test the
approximation; and we also thank D. Barbosa for supplying much information
concerning current experimental results. 
\end{acknowledgements}

\end{document}